# Nonclassical Nature of Dispersion Cancellation and Nonlocal Interferometry

J.D. Franson
*Physics Department, University of Maryland, Baltimore County, Baltimore, MD 21250*
*(27 July, 2009)*

Several recent papers have shown that some forms of dispersion cancellation have classical analogs and that some aspects of nonlocal two-photon interferometry are consistent with local realistic models. It is noted here that the classical analogs only apply to local dispersion cancellation experiments [A.M. Steinberg et al., Phys. Rev. Lett. **68**, 2421 (1992)] and that nonlocal dispersion cancellation [J.D. Franson, Phys. Rev. A **45**, 3126 (1992)] is inconsistent with any classical field theory and has no classical analog. The local models that have been suggested for two-photon interferometry are shown to be local but not realistic if the spatial extent of the interferometers is taken into account. It is the inability of classical models to describe all of the relevant aspects of these experiments that distinguishes between quantum and classical physics, which is also the case in Bell's inequality.

## I. INTRODUCTION

Energy-time entanglement gives rise to a number of phenomena that appear to be nonclassical and nonlocal, including nonlocal two-photon interferometry [1-3], dispersion cancellation [4-7], and nonlocal phase modulation [8-9]. Although it may seem apparent that these effects are nonclassical, it is important to understand the extent to which classical models can or cannot describe these phenomena. This provides additional insight into the nature of entanglement and it may be of practical benefit if classical effects can simulate some of the useful properties of these effects [10-13].

At the same time, it is equally important to ask whether or not classical models can describe all of the relevant aspects of a given experiment. Classical models may agree with some aspects of an experiment but not others. If it can be shown that no classical model can agree with all of the results of an experiment, then that phenomena must be considered to be nonclassical in nature. The same is true for local realistic theories that only agree with some aspects of nonlocal interferometer experiments.

For example, consider violations of Bell's inequality [14] based on the polarization entanglement of pairs of photons. It is straightforward to contrive local realistic models that agree with the results of those experiments if the polarization analyzers are always set to the same orientation. Bell's inequality depends on the fact that the experimenter can choose to make measurements at several different settings of the analyzers. It is the inability of hidden variable models to describe all of the relevant aspects of the experiments that distinguishes between quantum mechanics and local realism.

Several recent papers [10-13] have shown that classical fields can simulate the effects of dispersion cancellation in certain types of experiments. As a result, there has been a tendency to assume that all dispersion cancellation is classical in nature or has classical analogs. This situation is discussed in Section II, where it is pointed out that these classical models only apply to the local dispersion cancellation experiments of Steinberg et al. [6, 7], while nonlocal dispersion cancellation [4, 5] is inconsistent with classical field theory and has no classical analog. As a result, it is important to make a distinction between local and nonlocal dispersion cancellation when considering the classical or quantum nature of the phenomena. Section III shows that the nonlocal phase modulation recently proposed by Harris [8-9], which is a quantum analog of nonlocal dispersion cancellation, is also inconsistent with classical field theory.

The limited ability of classical models to describe two-photon interferometry is discussed in Section IV. Cabello et al. [15] recently proposed a local hidden variable model that is consistent with some aspects of nonlocal two-photon interferometry. It is shown that models of this kind are local but not realistic if the spatial extent of the interferometers is taken into account. In particular, models of that kind neglect the experimental fact that a single photon can only be detected in one path or the other of an interferometer. Once again, it is important to consider all aspects of a given experiment before concluding that the two-photon interferometer experiment of Ref. [1] cannot rule out local realistic theories.

A summary and conclusions are provided in Section V.

## II. DISPERSION CANCELLATION

The original proposal for dispersion cancellation [4] is illustrated in Fig. 1. A pair of energy-time entangled photons is generated in the source using parametric down-conversion, in which individual photons in a pump laser are split into pairs of photons, conserving energy in the process. Photon pairs generated in this way are emitted at very nearly the same time and each photon has a relatively large bandwidth. Nevertheless, the sum of their frequencies is equal to that of the pump laser in the down-conversion source and the two frequencies are anti-correlated. The two photons then propagate through two dispersive media that are separated by a large distance. It was shown in Ref. [4] that the dispersion in one medium can be cancelled out by the dispersion in the other medium if the dispersion coefficients have the opposite sign, and the two photons will still be detected at the same time even though they have both passed through a dispersive medium. This effect has been demonstrated experimentally using the dispersion of optical fibers [16] or gratings [17].

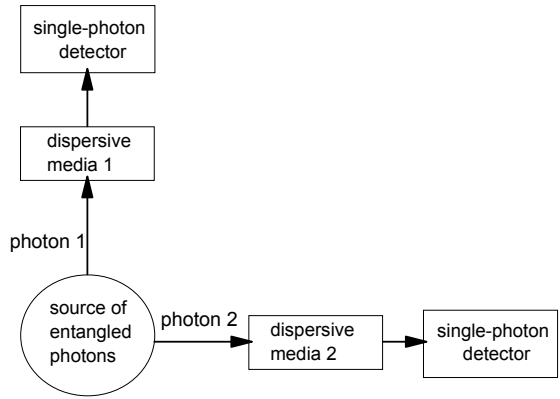

Fig. 1. Nonlocal dispersion cancellation [4], in which a pair of energy-time entangled photons propagate through two distant dispersive media. The dispersion in one medium can be cancelled out nonlocally by the dispersion in the other medium if the dispersion coefficients of the two media are equal and opposite.

Nonlocal dispersion cancellation can be understood by considering an initial entangled state of the form

$$|\psi\rangle = c_n \int_0^\infty d\omega_1 \left\{ Exp\left[-(\omega_1-\omega_0)^2/2\sigma_F\right]\right\}^2 \times \hat{a}_1^\dagger(k_1)\hat{a}_2^\dagger(2k_0-k_1)|0\rangle. \quad (1)$$

Here $c_n$ is a normalization constant, $\hat{a}_1^\dagger(k_1)$ creates a photon in path 1 with frequency $\omega_1$ and wave vector $k_1 = \omega_1/c$ (in free space), and $\hat{a}_2^\dagger(k_2)$ creates a photon with frequency $\omega_2$ in path 2 with $k_2 = \omega_2/c$ ($c$ is the speed of light). The frequency bandwidth $\sigma_F$ can be controlled by passing the photons through a filter, as is often done experimentally. The central frequency $\omega_0 = ck_0$ of the photons is half that of the pump laser, and we have used the fact that $k_2 = 2k_0 - k_1$ for photons that are anti-correlated in frequency.

The effects of the dispersive media on the photons can be investigated by writing the frequencies of the two photons in terms of a single parameter $\varepsilon$:

$$\omega_1 = \omega_0 + \varepsilon$$
$$\omega_2 = \omega_0 - \varepsilon. \quad (2)$$

Inside the media, the wave vector $k$ becomes a function of the frequency. To second order, we can expand $k(\omega)$ in the form

$$k_1 = k_0 + \alpha_1\varepsilon + \beta_1\varepsilon^2$$
$$k_2 = k_0 - \alpha_2\varepsilon + \beta_2\varepsilon^2 \quad (3)$$

where the coefficients $\alpha_i$ and $\beta_i$ are related to the group velocity and dispersion, respectively. It was shown in Ref. [4] that the standard deviation $\sigma_T$ of the difference in detection times of the photons is given by

$$\sigma_T^2 = \frac{1/\sigma_F^4 + (\beta_1+\beta_2)^2 L^2}{1/\sigma_F^2} \quad (4)$$

where $L$ is the length of the dispersive media. It can be seen that the dispersive media have no effect if $\beta_1 = -\beta_2$. The fact that these coefficients are added first and then squared is characteristic of a quantum interference effect. In this case, the destructive interference is possible because of the coherent nature of the integral over $\omega_1$ in Eq. (1).

The entangled state of Eq. (1) corresponds to the use of a CW laser with a well-defined frequency. If the pump laser produces short pulses instead, then the analysis is more complex and the effects of dispersion cancellation may be reduced or eliminated [18].

The corresponding calculation can be performed for two classical pulses with the same bandwidth $\sigma_F$ [4]. The standard deviation $\sigma_C$ of the difference in the detection times of the classical pulses (using single-photon detectors) is given instead by

$$\sigma_C^2 = \frac{1/(2\sigma_F^4) + (\beta_1^2+\beta_2^2)L^2}{1/\sigma_F^2}. \quad (5)$$

Here the dispersive coefficients $\beta_1$ and $\beta_2$ are squared first and then added, which is characteristic of the incoherent sums associated with classical probabilities.

The difference between classical dispersion and nonlocal dispersion cancellation can be illustrated by considering the following example, which has been suggested as a possible classical analog for nonlocal dispersion cancellation. Consider a classical source that contains two lasers that produce a sequence of short pulses with a frequency bandwidth $\sigma_P$, as illustrated in Fig. 2. The pulses will be labeled with an index $i$ and the central frequencies $\bar{\omega}_{1i}$ and $\bar{\omega}_{2i}$ of the individual pulses will be assumed to be anti-correlated:

$$\bar{\omega}_{1i} = \omega_0 + \delta\omega_i$$
$$\bar{\omega}_{2i} = \omega_0 - \delta\omega_i. \qquad (6)$$

Here $\omega_0$ is a constant frequency and $\delta\omega_i$ is assumed to vary randomly from one pulse to the next as described by a probability distribution $P(\delta\omega_i)$, such as

$$P(\delta\omega_i) = c_{n'} Exp[-(\delta\omega_i)^2 / 2\sigma_D^2]. \qquad (7)$$

Here $c_{n'}$ is a normalization constant and $\sigma_D$ describes the overall bandwidth of the source, as would be observed in the time-averaged spectra of the two beams.

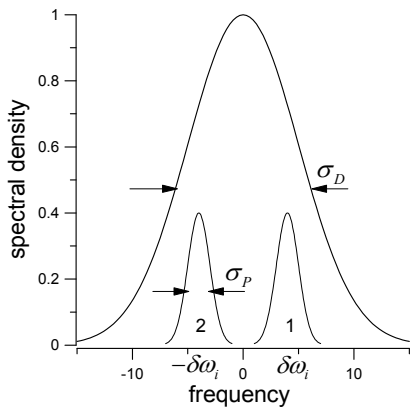

Fig. 2. Classical model in which the source consists of two lasers that emit a sequence of short pulses whose frequencies are anti-correlated (arbitrary units). The frequency of pulse 1 is shifted by a random amount $\delta\omega_i$ from the time-averaged frequency, while the frequency of pulse 2 is shifted by $-\delta\omega_i$. The overall (time-averaged) bandwidth $\sigma_D$ of the source is determined by the probability distribution for $\delta\omega_i$. This model gives spatial correlations between the intensities of the two beams after they pass through two dispersive media, but dispersion still occurs and this model is not analogous to nonlocal dispersion cancellation.

When these classical pulses pass through two dispersive media, the length of each individual pulse will be increased by dispersion. But their intensities at the detectors will still be correlated to some extent if $\beta_1 = -\beta_2$. For example, a positive value of $\delta\omega_i$ will cause the pulse in the medium with the normal dispersion coefficient to propagate more slowly than average, while the corresponding negative value of $\delta\omega_i$ in the medium with the anomalous dispersion coefficient will also cause that pulse to propagate more slowly. Thus the group velocities are the same in both media and some degree of spatial correlation will be maintained between the two pulses.

Although this example does show classical correlations between the intensities of the two pulses, it is not analogous to nonlocal dispersion cancellation for several reasons. For each set of classical pulses labeled with index $i$, the dispersive coefficients $\beta_1$ and $\beta_2$ must still be squared and then summed in Eq. (5), which indicates that there is no coherent cancellation of probability amplitudes as there is in Eq. (4) for the quantum-mechanical case. The length $l_C$ of the classical pulses is initially $\sim c/\sigma_P$ and this is increased further by dispersion. In constrast, energy-time entangled photons with the same bandwidth $\sigma_D$ as the classical source will be correlated in position to within an uncertainty $l_Q \sim c/\sigma_D$, with $l_Q \ll l_C$. It is not difficult to prove in general that there is no classical model for which the pulses remain correlated to within the inverse of the total bandwidth of the source, both in free space and when the dispersive media are inserted, as is the case quantum-mechanically.

Torres-Company et al. [19] have discussed another example in which they showed that classical stationary chaotic light (e.g., white light) maintains the same fourth-order interference properties after two identical beams have passed through two identical dispersive media. In this case, a chaotic beam of light was assumed to pass through a 50/50 beam splitter and intensity correlations between the reflected and transmitted beams were assumed to be measured using Hanburry-Brown and Twiss interferometry, as illustrated in Fig. 3. The usual factor of two peak in the intensity correlation function was found to be unaltered if identical dispersive media are placed in the paths of the two beams, which Torres-Company et al. interpreted as showing that "nonlocal dispersion cancellation has a classical counterpart".

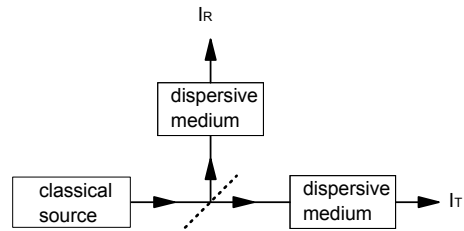

Fig. 3. A classical source of chaotic light is split by a 50/50 beam splitter, as suggested by Torres-Company et. al [19]. The reflected and transmitted beams, which are identical, are passed through two identical dispersive media, and the correlations between the intensities of the two beams are measured (Hanburry-Brown and Twiss interferometry).

This effect had already been discussed in the first paper on dispersion cancellation [4], where it was noted that no dispersion cancellation at all occurs in such an effect involving chaotic light. What actually happens is that both beams undergo identical dispersion in which the location of the various peaks and dips in the intensity are changed in the same way, so that the two beams remain identical, as illustrated qualitatively in Fig. 4. It can be shown [20] that dispersion does not alter the statistical moments of a chaotic field, and thus the intensity correlations between the two beams remain the same. It seems apparent here that the dispersion of each beam is the result of a local process and that there is no dispersion cancellation or other nonlocal interaction between the two beams, unlike the situation of Fig. 1. Furthermore, single-photon detectors would register counts at widely different times and the situation is not analogous in any way to dispersion cancellation. It would be more correct to refer to effects of this kind as "identical dispersion" rather than dispersion cancellation. Similar comments apply to a classical model recently proposed by J.H. Shapiro [21].

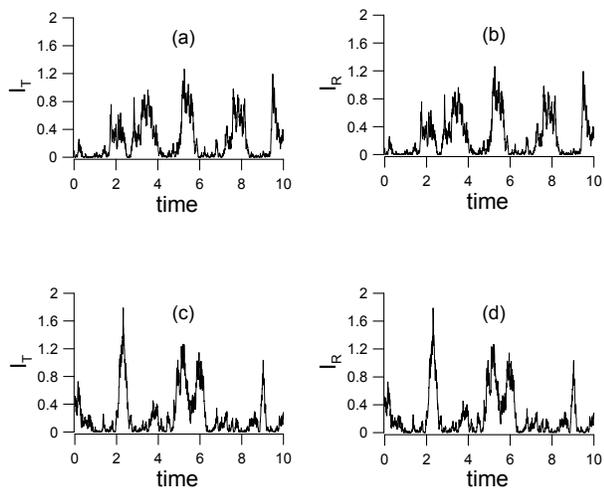

Fig. 4. Typical intensities of a classical beam of chaotic light passed through a beam splitter as suggested by Torres-Company et al. [19], plotted as a function of time (arbitrary units). (a) Intensity of the transmitted beam of light with no dispersive medium. (b) Corresponding intensity of the reflected beam of light with no dispersive medium. (c) Intensity of the transmitted beam of light after passing through a dispersive medium. (d) Corresponding intensity of the reflected beam of light after passing through an identical dispersive medium. There is no cancellation of dispersion in this example; instead, both beams are dispersed in the same way, which maintains the statistical correlations between the intensities of the two beams [4, 20]. (These plots were simulated in Mathematica using a Markov process.)

Steinberg et al. [6, 7] independently proposed and demonstrated a different kind of dispersion cancellation as illustrated in Fig. 5. Here two energy-time entangled photons propagate along the two arms of a Hong-Ou-Mandel interferometer [22], with a dispersive element in one of the two paths. The two beams are brought back together to interfere at a single beam splitter. Steinberg et al. showed that the effects of dispersion will cancel out in the observed coincidence rates in an interferometer of this kind. It is important to note, however, that the interference occurs after the beams have been recombined at a single location, that of the final beam splitter, and that the interference is a local process as a result. In order to distinguish this form of dispersion cancellation from the nonlocal dispersion cancellation of Fig. 1, it will be referred to here as local dispersion cancellation. The possibility of classical analogs for local dispersion cancellation of this kind cannot be ruled out [23].

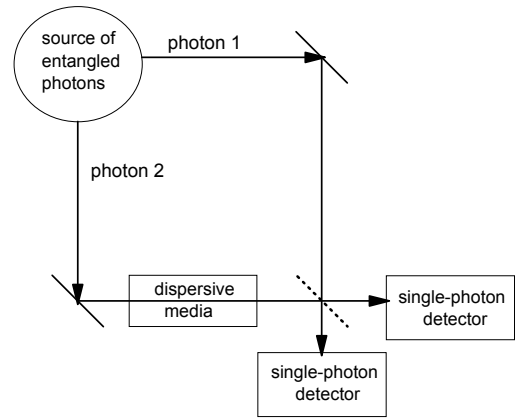

Fig. 5. Local dispersion cancellation as proposed and demonstrated by Steinberg et al [6, 7]. A pair of energy-time entangled photons propagate through the two arms of a Hong-Ou-Mandel interferometer [22], with a dispersive media present in one of the paths. Steinberg et al. showed that the width of the well-known dip in the coincidence rate was unaffected by the dispersive media. Here the photons do not interfere until they have been recombined at the beam splitter, and the local nature of this interference allows classical models to give analogous results.

Resch et al. [10] have demonstrated local dispersion cancellation using the same general experimental setup of Fig. 5 but with the entangled photons replaced with a classical source of light (a broad-band laser). The output of their experiment contained a term that is equivalent to the dispersion cancellation in the quantum-mechanical case, but with a visibility that is reduced by 50%. This reduction in the visibility is typical of classical models for two-photon interferometers, as will be discussed further in the next Section.

Subsequent experiments by Kaltenbaek et al. [11] also demonstrated local dispersion cancellation for classical beams of light but with visibilities > 85%. The experimental geometry was once again similar to that of Fig. 5, but in this case chirped and anti-chirped laser pulses with anti-correlated frequencies were transmitted through the two arms of the interferometer. A nonlinear crystal was used to perform sum-frequency generation, which produce an output beam with frequency $\omega_S = \omega_1 + \omega_2$. This allows the dispersion to be cancelled with visibilities that can approach 100% in the ideal case. The local nature of this process is apparent from the fact that the two beams must be brought back together at the same location (that of the nonlinear crystal) in order to obtain the dispersion cancellation in the output.

Although the quantum interference occurs locally in the dispersion cancellation experiment of Fig. 5 [6, 7], that does not mean that those effects do not depend on the entanglement of the photons. The classical experiments were not performed in the same way as in Fig. 5 and it seems unlikely that any classical theory could completely reproduce the results of local dispersion cancellation experiments [24].

Refs. [12] and [13] discuss further classical analogs of effects that are closely related to local dispersion cancellation. These effects may be of practical use in optical coherence tomography or imaging applications [18], and they provide further insight into the nature of these effects. But all of these classical analogs pertain to the local dispersion cancellation [6, 7] of Fig. 5 or related effects and not to the nonlocal dispersion cancellation [4, 5] of Fig. 1. As a result, they do not provide any evidence for a classical analog of nonlocal dispersion cancellation.

Ref. [4] showed that nonlocal dispersion cancellation is inconsistent with classical theories and it does not have a classical analog; the coherent cancellation of quantum-mechanical probability amplitudes at distant locations that is evident in Eq. (4) cannot be reproduced by incoherent sums over classical probabilities, as can be seen from Eq. (5).

### III. NONLOCAL PHASE MODULATION

Harris has recently proposed [8, 9] a new effect that is a quantum analog of nonlocal dispersion cancellation, which he refers to as nonlocal modulation. As illustrated in Fig. 6, pairs of energy-time entangled photons are produced by parametric down-conversion and propagate towards two distant phase modulators that operate at a fixed frequency $\Omega$. After the photons pass through the phase modulators, their frequency spectrum is measured using identical monochromators and the correlations between their frequency spectra are recorded.

If a classical monochromatic field at frequency $\omega_C$ is passed through such a phase modulator, it will produce sidebands at frequencies of $\omega_C \pm n\Omega$, where $n$ is an integer. This would reduce the frequency anti-correlation of two classical beams of light, since the production of the sidebands occurs independently at the two locations. Harris [8] showed that the energy-time entangled photons from down-conversion will remain anti-correlated in frequency despite the modulators, provided that the phase shift $\phi_1(t)$ produce by one of the modulators is equal and opposite to the phase shift $\phi_2(t)$ produced by the other modulator. This effect was recently demonstrated experimentally by Sensarn et al. [9].

The nonclassical nature of nonlocal modulation of this kind can be shown as follows. Suppose that we generate classical pulses of light labeled by an index $i$ with probability $P_i$. For each pulse $i$, the electric fields associated with beams 1 and 2 must have some value $E_{1i}(\mathbf{r},t)$ and $E_{2i}(\mathbf{r},t)$ in any classical theory [2]. For a specific pulse, those fields must also have some Fourier transform $\mathcal{E}_{1i}(\omega_1)$ and $\mathcal{E}_{2i}(\omega_2)$ with corresponding spectral densities $\mathcal{P}_{1i}(\omega_1)$ and $\mathcal{P}_{2i}(\omega_2)$. As a measure of the degree of anti-correlation between the two classical beams after they have passed through the modulators, we can consider the standard deviation $\Delta$ of the sum of the measured frequencies $(\omega_1 + \omega_2)$ given by

$$\Delta^2 = c_n" \sum_i P_i \iint d\omega_1 d\omega_2 \mathcal{P}_{1i}(\omega_1)\mathcal{P}_{2i}(\omega_2) \times \left[(\omega_1 + \omega_2) - 2\bar{\omega}\right]^2. \tag{8}$$

Here $2\bar{\omega}$ is the average value of $\omega_1 + \omega_2$ and $c_n"$ is a normalization constant required to convert the spectral densities into probability densities. If the two beams are completely anti-correlated, then $\Delta$ should be zero.

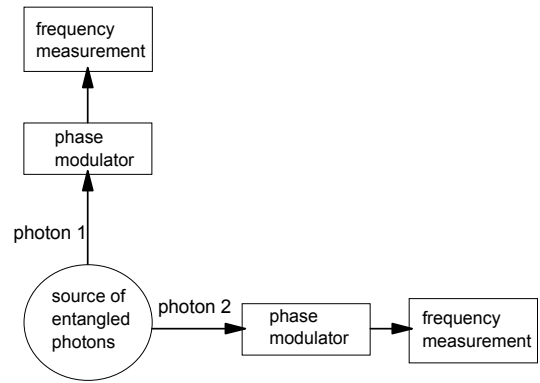

Fig. 6. Nonlocal phase modulation proposed by Harris [8]. A pair of energy-time entangled photons propagate along different paths where they pass through identical phase modulators. The correlations between the frequencies of the two photons are then measured. The effects of one phase modulator are cancelled out by the other, distant modulator if the two modulators produce equal but opposite phase shifts as a function of time.

With the phase modulators turned off, the two beams of light must be anti-correlated to within a small frequency uncertainty in order to be consistent with the quantum-mechanical predictions. For simplicity, we will first consider the case in which the two frequencies are totally anti-correlated and return to the more general case later. In that case we could measure the frequency of beam 1 and predict what the frequency of beam 2 must be. That can only be the case if beam 2 corresponds to a monochromatic beam of light with the predicted frequency, since the measurement of the frequency of beam 1 cannot change the pulse sent into path 2 (the

usual local realism argument). In that case, we must have that

$$\mathcal{P}_{1i}(\omega_1) = a_i \delta(\omega_1 - \omega_i)$$
$$\mathcal{P}_{2i}(\omega_2) = b_i \delta(\omega_2 + \omega_i - 2\bar{\omega}). \quad (9)$$

Here $a_i$ and $b_i$ are positive constants for any given pulse (related to their energies) and $\omega_i$ is the frequency that would be measured for beam 1 for that particular pulse.

In any classical theory, the nature of the output pulses cannot depend on the choice of whether or not we decide to turn on the modulators. As a result, Eq. (9) must still hold for the fields emitted by the source whether the modulators are turned on or not. The response of a monochromatic classical field to a phase modulator is well known and consists of the sidebands mentioned above, so that after the beams have passed through the modulators the spectral densities become

$$\mathcal{P}_{1i}'(\omega_1) = a_i \sum_{n=0}^{\infty} f_n \delta(\omega_1 - \omega_i \pm n\Omega)$$
$$\mathcal{P}_{2i}'(\omega_2) = b_i \sum_{n=0}^{\infty} f_n \delta(\omega_2 + \omega_i - 2\bar{\omega} \pm n\Omega). \quad (10)$$

Here the $\pm$ sign indicates that we include terms with both signs in the sum and the $f_n$ are positive coefficients related to the squares of Bessel functions of integer order, whose values depend on the amplitude of the phase modulation and are of no interest here.

Having deduced the form of the power spectral densities in the classical case, we can now calculate the value of $\Delta$:

$$\Delta^2 = c_n'' \sum_i P_i \iint d\omega_1 d\omega_2 a_i \sum_n f_n \delta(\omega_1 - \omega_i \pm n\Omega)$$
$$\times b_i \sum_{n'} f_{n'} \delta(\omega_2 + \omega_i - 2\bar{\omega} \pm n'\Omega) \left[(\omega_1 + \omega_2) - 2\bar{\omega}\right]^2. \quad (11)$$

Using the $\delta$ functions, this expression reduces to

$$\Delta^2 = 2 c_n'' \Omega^2 \sum_i P_i a_i b_i \sum_{n,n'} f_n f_{n'}$$
$$\times \left[(n+n')^2 + (n-n')^2\right]. \quad (12)$$

Once again, $P_i$, $c_n''$, $a_i$, $b_i$, $f_n$ and $f_{n'}$ are all positive constants that depend on the amplitude of the phase modulation and the average intensities of the two beams, so that all of the terms in the sum are positive and $\Delta^2$ is nonzero. Eq. (12) shows that the effects of the phase modulators cannot cancel out for classical light, unlike the case for energy-time entangled photons. In particular, Eq. (12) does not depend on the relative phase of the two distant modulators.

Eq. (9) corresponds to the limiting case in which the two beams are totally anti-correlated. If the beams are only anti-correlated to within a frequency uncertainty $\delta\omega$, then the Dirac $\delta$-functions in Eq. (9) would have to be replaced with narrow spectral distributions $d(\omega)$ with widths on the order of $\delta\omega$. It is not difficult to show that the results of Eq. (12) will still be obtained in this case as long as $\delta\omega$ is sufficiently small compared with $\Omega$ that the sidebands do not overlap.

Eq. (12) shows that there is no classical model that is consistent with nonlocal phase modulation, but it is still possible to consider classical models that are in agreement with some aspects of the experiment but not others. For example, suppose that we chose to neglect the fact that the frequencies are anti-correlated when the phase modulators are turned off. In that case, we could construct a source that gives complete anti-correlation when the phase modulators are turned on but not when they are turned off. This can be done by arranging for the source to first produce two laser pulses whose frequencies are well-defined during a given pulse and anti-correlated about a central frequency, as in Eq (9). In addition, phase modulators can be placed inside the source to produce a phase shift of $-\phi_1(t)$ and $-\phi_2(t)$ on the two beams before they leave the source. When the beams pass through the distant phase modulators, those modulators will apply phase shifts that cancel the ones applied inside the source, and the final two beams will have completely anti-correlated frequencies, in agreement with that aspect of the experiment.

This example illustrates the need to consider classical models that agree with all aspects of the experiment in question and not just some of its features. Is it reasonable to say that nonlocal phase modulation has a classical analog since we can construct classical models of the kind discussed above? The limitations of this model (and others like it) show that there really is no classical analog for these effects. Just as is the case for Bell's inequality, it is the inability of classical models to describe all of the relevant aspects of an experiment that allows us to distinguish between quantum and classical physics.

## IV. NONLOCAL INTERFEROMETRY

Energy-time entanglement allows nonlocal interference experiments [1-3] that violate Bell's inequality, as illustrated in Fig. 7. In these experiments, two identical interferometers have a short path S and a longer path L. Pairs of photons in an energy-time entangled state are incident on the two interferometers. Correlations are observed between those pairs of photons that arrive at the detectors at the same time (coincident detection), which means that both photons must have taken the longer paths (LL events) or both must have taken the shorter paths (SS events). Those correlations were shown [1] to violate Bell's inequality, but their interpretation is complicated by the fact that only the coincident events corresponding to LL or SS are accepted in a post-selection process. The predicted effects were initially met with skepticism, but they have now been verified experimentally [3, 25-31].

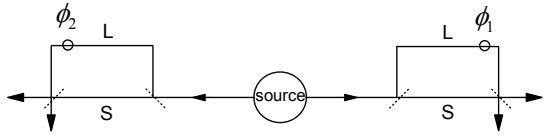

Fig. 7. A pair of energy-time entangled photons incident on two distant interferometers with a shorter path S and a longer path L [1]. Phase shifts $\phi_1$ and $\phi_2$ are inserted into the two longer paths. The output of this interferometer has been shown to be in disagreement with any local realistic theory.

Ou and Mandel [32] considered a classical model in which the source consisted of two lasers whose frequencies were anti-correlated and showed that the output of the interferometers would have the same phase dependence as in the quantum-mechanical case but with a maximum visibility of 50%. Since a visibility of 71% is required to violate Bell's inequality, this model did not show that the interferometer of Fig. 7 is consistent with local realism. This classical model also does not describe the fact that the photons are emitted at the same time and that the detection events will be coincident to within a small timing uncertainty that is inversely proportional to the bandwidth of the photons as in Section II. This situation was used in Ref. [2] to derive an inequality that must be satisfied by any classical field theory:

$$v \leq \frac{R_{C0}(\Delta t)}{R_{C0}(0) + R_{C0}(\Delta t)}. \quad (13)$$

Here $R_{C0}(\Delta t)$ is the coincidence counting rate that would be obtained with no interferometers in place for a time offset of $\Delta t$ and $v$ is the visibility of the two-photon interference fringes. This inequality can be violated for visibilities that are much less than 50%, as has been demonstrated experimentally [3], and it contradicts the commonly-held belief that classical theories can be consistent with the quantum-theory prediction for experiments of this kind with visibilities up to 50%.

Aerts et al. [33] showed that there are local hidden variable models that agree with the output of the interferometer if the interferometer is taken to be a "black box" and its internal structure is ignored. This is because the post-selection process is analogous to the usual detection loophole, and Bell's inequality can be violated by classical models if the fair-sampling assumption is not made. Aerts et al. [33] went on to show that the results of the experiment are inconsistent with any local realistic theory, however, if the temporal delays of the interferometer are taken into account. Their proof was based on the use of chained Bell inequalities and it requires measurements to be made at different phase settings from the usual CHSH inequality.

The role of post-selection in these experiments can be further understood [34] by noting that the presence or absence of a single photon in either path through an interferometer can be predicted with certainty by placing a detector in the other path of the same interferometer, as has been demonstrated in numerous experiments. The detection process can be sufficiently fast that it is space-like separated from the other path through the same interferometer, so that the measurement cannot disturb the state of the field in the other path. (This is not difficult to accomplish in actual experiments.) As a result, the path taken by a photon is an element of reality as defined by Einstein, Podolsky, and Rosen (EPR) [35] that must be determined by any local realistic theory while the photons are still traversing the interferometers. The LS and SL events do not contribute to the statistics collected and the remaining LL and SS events are described by a reduced set of hidden variables that must determine the results of the subsequent measurements without any further post-selection. As a result, it was shown in Ref. [34] that the interferometer of Fig. 7 is inconsistent with any local hidden-variable theory, even when using the original CHSH inequality and including the effects of post-selection.

Cabello et al. [15] recently proposed another local model that can duplicate some aspects of these experiments. In their model, the S/L decision "may be made as late as the detection time $t_D$" after the photons have already traversed the interferometer. But the presence or absence of a photon in either path can be predicted with certainty while the photons are still traversing the interferometer and this is an element of reality that must be specified in any local realistic theory. Cabello et al. explain this aspect of their model by noting that "the notion of EPR elements of reality was abandoned." As a result, their model is not a local realistic theory as defined by EPR [35]; it is local but not realistic. Models of this kind have already been criticized for being inconsistent with other experiments [36].

One could argue that the model of Cabello et al. was intended to treat the interferometers as "black boxes" and to ignore any physics that may take place inside the interferometers themselves. As we have seen in several examples earlier in this paper, it is not difficult to find classical models that agree with some aspects of an experiment while disagreeing with other relevant aspects of the experiment. In this case, their model ignores the fact that a single photon can only be detected in one location. If their model does not determine the path taken by a photon while it is still traversing one of the interferometers, then it would be possible to detect a single photon more than once. There may be some merit in considering models of this kind, but it does not seem reasonable to conclude that the experiment of Ref. [1]

cannot rule out local realism, as Cabello et al. stated, unless their model is consistent with all relevant aspects of the experiment.

It is true that no nonlocal interferometer experiments of this kind have been performed to date in which a detector could be rapidly inserted into the paths of the interferometers to verify that a single photon can only be detected in one path but not both. But numerous experiments have already demonstrated this property using a single photon and a beam splitter. By its very nature, realism assumes that nature has this property regardless of whether or not we choose to measure it. As a result, there is no need to repeat this demonstration in every subsequent test of local realism; we already know the results of such a measurement, and in a realistic theory those properties must hold regardless of whether or not we choose to measure them.

It has been suggested that the proof of Ref. [34] cannot be applied to the model of Cabello et al. on the grounds that Ref. [34] assumed rapidly-varying phase shifters, whereas the model of Cabello et al. was not intended to apply to that situation. Rapidly varying settings of the measurement devices are required in all experimental tests of Bell's inequality in order to rule out the possibility of signals traveling from the measurement devices to the source at the speed of light, however unlikely that may seem. This is true for the Bell inequality experiments based on polarization entanglement that were performed by Aspect et al. [37], for example, and this requirement is not unique to the experiment of Fig. 7. In particular, the modified interferometer suggested by Cabello et al. [15] would also require rapidly-varying phase settings to rule out such a possibility.

The EPR definition of an element of reality (or local realism) states that "If, without in any way disturbing a system, we can predict with certainty the value of a physical quantity, then there exists an element of physical reality corresponding to this physical quantity". Larsson [38] has recently suggested that the use of EPR elements of reality as described above may not be appropriate, since a measurement of the presence or absence of a photon in one path would disturb the interference pattern produced by the interferometer. One could still assume that the path must be specified in any local realistic theory, but the question is whether or not this follows unavoidably from the definition of a local realistic theory or if it need not hold in general.

Larson's comment raises the interesting point that the EPR definition of an element of reality is dependent on the sometimes arbitrary division of nature into separate "systems". If it is apparent that two systems are totally isolated, then the usual EPR definition is adequate, as illustrated in Fig. 8a. But in the situation of interest here, we initially have two separated systems consisting of the fields in the two distant paths through the interferometer. After the photons reach the second beam splitter, those systems interact and form another system of interest corresponding to the output of the interferometer, as illustrated in Fig. 8b.

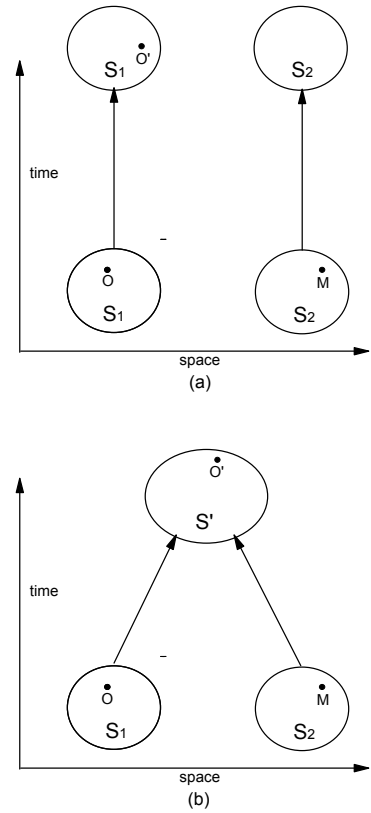

Fig. 8. Space-time diagrams illustrating the definition of an element of reality (arbitrary units). (a) Here the system of interest can be divided into two separate systems $S_1$ and $S_2$. If the value of an observable property $O$ in system $S_1$ can be predicted with certainty by making a measurement $M$ at a space-like separated point in system $S_2$, then $O$ is said to be an element of reality and any local realistic theory must determine its value at that time. (b) Here the two systems $S_1$ and $S_2$ are initially separated, but they later merge to form a new system $S'$. Once again, if the value of an observable property $O$ can be predicted with certainty by making a measurement $M$ at a space-like separated point, then $O$ is said to be an element of reality and any local realistic theory must determine its value at that time. The value of $O$ must be determined by any local realistic theory even if we choose not to make measurement $M$, and this fact can subsequently be used to make inferences about the predictions of any local realistic theory for another observable $O'$ even if $O'$ is in the forward light cone of $M$.

This suggests that a more general definition of an element of reality that does not require the division of the system of interest into subsystems may be appropriate: "If the value of an observable property $O$ can be predicted with certainty at location **x** and time $t$ by making a space-like separated measurement $M$ at another location **x'** and time $t'$, then the property $O$ is defined to be an element of reality at that time." A local realistic theory must determine all such elements of reality in order to be consistent with the usual definition of realism. Although this definition is somewhat more general than the original EPR definition, it is consistent with causality and their intended meaning of an element of reality.

It is important to note that the value of $O$ must be determined by any local realistic theory even if we choose not to make measurement $M$, and that fact can be used to make inferences regarding the possible predictions of any local realistic theory for another observable $O'$ even if $O'$ is in the forward light cone of $M$.

Returning to the proof in Ref. [34], any local realistic theory must determine the paths through the interferometer immediately after the photons leave the beam splitters, since a measurement $M$ in one path through an interferometer could be used to predict with certainty the presence or absence of the photon in the other, space-like separated path of the same interferometer. The paths must be determined regardless of whether or not such a measurement is actually made, and the interferometers are clearly not disturbed if no such measurement is made. In that case, the fact that the paths must have been chosen anyway can be used to limit the predictions of any hidden-variable theory regarding the output $O'$ of the interferometers, as was done in Ref. [34]. This argument shows that the model of Cabello et al. [15] is local but not realistic, and that there is no local realistic model that is consistent with the predictions of quantum mechanics for the interferometer of Fig. 7.

To summarize this section, we see once again that there are classical models that are consistent with some aspects of nonlocal two-photon interferometer experiments. But there are other essential aspects of the experiments that are not described by these models, such as the inability to detect a single photon in both paths of an interferometer. The connection between this recurring feature of the proposed classical models and wave-particle duality will be considered in the next section.

## V. SUMMARY AND CONCLUSIONS

It is always important to consider the extent to which quantum effects can be reproduced by classical models. There are examples of effects that were originally believed to be inherently quantum-mechanical (Hanburry-Brown and Twiss interferometry, for example) but were later found to be totally consistent with classical electromagnetism. Even when that is not the case, it is instructive to consider the possibility of classical analogs for quantum effects, and in some cases the classical analogs may be of practical use [10-13].

It is equally important to point out the ways in which a classical model does not completely describe a quantum effect of interest. If that is not done, we may incorrectly assume that a quantum effect is classical in nature when it is not.

In the local dispersion cancellation proposed by Steinberg et al. [6, 7], two photons are recombined on a single beam splitter and the local nature of the interference allows classical analogs. In this case there can be a coherent cancellation of classical field amplitudes that really is analogous to the cancellation of quantum-mechanical probability amplitudes, as has been demonstrated using sum-frequency generation [11]. That is not the case for the nonlocal dispersion cancellation [4, 5] of Fig. 1, where the photons never return to a common location and the quantum interference is truly nonlocal. There are no classical models that are consistent with nonlocal dispersion cancellation, and the models that have been suggested are qualitatively and quantitatively different. None of those models can mimic the cancellation of probability amplitudes at two distant locations that is responsible for nonlocal dispersion cancellation.

The nonlocal phase modulation proposed by Harris [8, 9] is a quantum-mechanical analog of nonlocal dispersion cancellation. It was shown here that there is no classical model that can reproduce the frequency anti-correlations that are predicted by the quantum theory with the modulator turned on and with it turned off. There are classical models that can maintain the anti-correlation for the case in which the modulators are turned on, but in that case the same model cannot maintain the anti-correlation if the modulators are turned off.

Local hidden variable models have been proposed that are capable of describing some aspects of nonlocal two-photon interferometry. Once again, those models ignore other relevant aspects of the experiments, such as the fact that the photons are coincident or the fact that a single photon can only be detected in one path through an interferometer. When all aspects of the experiments are taken into account, it can be shown that no local realistic theory is consistent with nonlocal interferometry [33, 34].

In many of these examples, there are classical models that can explain some aspects of an experiment but not all of its relevant features. This situation is analogous to the fundamental role of wave-particle duality in elementary quantum mechanics. Classical electromagnetism can describe the wave-like aspect of light, while Newton's corpuscular theory gives a qualitative description of its particle-like nature. Neither classical theory can explain both aspects of light. Does the fact that classical electromagnetism can describe the wave-like properties of light mean that all of quantum optics is classical in nature? Obviously not, and none of the classical models discussed here provide any evidence that nonlocal dispersion cancellation or nonlocal interferometry are classical in nature or consistent with local realism.

## ACKNOWLEDGEMENTS

I would like to acknowledge discussions with A. Cabello, J.-A. Larsson, T.B. Pittman, K. Resch, and A.

Steinberg. This work was supported in part by the National Science Foundation (NSF) under grant 0652560.

1. J.D. Franson, Phys. Rev. Lett. **62**, 2205 (1989).
2. J.D. Franson, Phys. Rev. Lett. **67**, 290 (1991).
3. J.D. Franson, Phys. Rev. A **44**, 4552 (1991).
4. J.D. Franson, Phys. Rev. A **45**, 3126 (1992).
5. M.J. Fitch and J.D. Franson, Phys. Rev. A **65**, 053809 (2002).
6. A.M. Steinberg, P.G. Kwiat, and R.Y. Chiao, Phys. Rev. Lett. **68**, 2421 (1992).
7. A.M. Steinberg, P.G. Kwiat, and R.Y. Chiao, Phys. Rev. A. **45**, 6659 (1992).
8. S.E. Harris, Phys. Rev. A **78**, 021807 (R) (2008).
9. S. Sensarn, G.Y. Yin, and S.E. Harris, submitted to Phys. Rev. Lett.
10. K.J. Resch, P. Puvanathasan, J.S. Lundeen, M.W. Mitchell, and K. Bizheva, Optics Express **15**, 8797 (2007).
11. R. Kaltenbaek, J. Lavoie, D.N. Biggerstaff, and K.J. Resch, Nature Physics **4**, 864 (2008).
12. J. Lavoie, R. Kaltenbaek, and K.J. Resch, Optics Express **17**, 3818 (2009).
13. R. Kaltenbaek, J. Lavoie, and K.J. Resch, Phys. Rev. Lett. **102**, 243601 (2009).
14. J.S. Bell, Physics (Long Island City, N.Y.) **1**, 195 (1964).
15. A. Cabello, A. Rossi, G. Vallone, F. De Martini, and P. Mataloni, Phys. Rev. Lett. **102**, 040401 (2009).
16. J. Brendel, H. Zbinden, and N. Gisin, Opt. Commun. **151**, 35 (1998).
17. S.-Y. Baek, Y.-W. Cho, and Y.-H. Kim, preprint arXiv:0811.2035 (2008).
18. J. Perina, Jr., A. V. Sergienko, B.M. Jost, B.E.A. Saleh, and M.C. Teich, Phys. Rev. A **59**, 2359 (1999).
19. V. Torres-Company, H. Lajunen, and A.T. Friberg, New J. Phys. **11**, 063041 (2009).
20. L.J. Wang, B.E. Magill, and L. Mandel, J. Opt. Soc. Am. B **6**, 964 (1989).
21. J.H. Shapiro, private communication.
22. C.K. Hong, Z.Y. Ou, and L. Mandel, Phys. Rev. Lett. **59**, 2044 (1987).
23. B.I. Erkmen and J.H. Shapiro, Phys. Rev. A **74**, 041601 (2006).
24. J.H. Shapiro and K.-X. Sun, J. Opt. Soc. Am. B **11**, 1130 (1994).
25. P.G. Kwiat, W.A. Vareka, C.K. Hong, H. Nathel, and R.Y. Chiao, Phys. Rev. A **41**, 2910 (1990).
26. Z.Y. Ou, X.Y. Zou, L.J. Wang, and L. Mandel, Phys. Rev. Lett. **65**, 321 (1990).
27. J. Brendel, E. Mohler, and W. Martienssen, Phys. Rev. Lett. **66**, 1142 (1991).
28. J.G. Rarity and P.R. Tapster, Phys. Rev. A **45**, 2052 (1992).
29. Y.H. Shih, A.V. Sergienko, and M.H. Rubin, Phys. Rev. A **47**, 1288 (1993).
30. D. Salart, A. Baas, C. Branciard, N. Gisin and H. Zbinden, Nature **454**, 861 (2008).
31. I. A. Khan and J. C. Howell, Phys. Rev. A **73**, 031801 (2006).
32. Z.Y. Ou and L. Mandel, J. Opt. Soc. Am. B **7**, 2127 (1990).
33. S. Aerts, P. Kwiat, J.-A. Larsson, and M. Zukowski, Phys. Rev. Lett. **83**, 2872 (1999).
34. J.D. Franson, Phys. Rev. A **61**, 012105 (1999).
35. A. Einstein, B. Podolsky, and N. Rosen, Phys. Rev. **47**, 777 (1935).
36. L.C. Ryff, Phys. Rev. Lett. **86**, 1908 (2001); preprint arXiv:0810.4825.
37. A. Aspect, J. Dalibard, and G. Roger, Phys. Rev. Lett. **49**, 1804 (1981).
38. J.-A. Larsson, to appear.